\documentclass{article}
\usepackage{amsmath}
\usepackage{amsfonts,amssymb,amsbsy,latexsym,amsthm}
\usepackage{graphicx}
\usepackage{subfigure}
\usepackage{authblk}
\usepackage{fullpage}

\usepackage{times}
\usepackage{bm}
\usepackage{natbib}

\def\ess{\textsc{ess}}

\newcommand{\tilX}{\widetilde{X}}

\newcommand{\bR}{\mathbb{R}}

\newcommand{\sX}{\mathsf{X}}

\newcommand{\sW}{\mathsf{W}}

\newcommand{\pitil}{\tilde{\pi}}

\newcommand{\pihat}{\hat{\pi}}

\newcommand{\Pbar}{\overline{P}}

\newcommand{\Vars}[2]{\mathrm{var}_{#1}\left(#2\right)}
\newcommand{\Var}[1]{\mathrm{var} \BK{ #1 }}

\newcommand{\Expect}[1]{\mathrm{E}\left(#1\right)}
\newcommand{\Prob}[1]{\mathrm{pr}\left({#1}\right)}

\newcommand{\md}{{\rm d}}

\newcommand{\ulw}{\underline{w}}  
\newcommand{\ulW}{\underline{W}}  
\newcommand{\tilPPM}{\widetilde{\mathsf{P}}}     
\newcommand{\PPM}{\mathsf{P}}     

\newcommand{\Dirichlet}{\mathcal{E}}

\newcommand{\lVar}{\mathrm{V}} 
\newcommand{\lVarCont}{\widetilde{\lVar}} 

\renewcommand{\phi}{\varphi}
\renewcommand{\epsilon}{\varepsilon}
\renewcommand{\rho}{\varrho}

\newcommand{\BK}[1]{ {\left( #1 \right)} }       
\newcommand{\sqBK}[1]{ {\left[ #1 \right]} }      
\newcommand{\curBK}[1]{ {\left\{ #1 \right\}} }

\newcommand{\norm}[1]{\left\Vert #1 \right\Vert}

\renewcommand{\tilde}{\widetilde}
\renewcommand{\hat}{\widehat}

\newtheorem{theorem}{Theorem}
\newtheorem{remark}{Remark}
\newtheorem{corollary}{Corollary}
\newtheorem{proposition}{Proposition}

\title{Pseudo-marginal Metropolis--Hastings using averages of unbiased estimators}

\author[1]{Chris Sherlock\thanks{anthony.lee@warwick.ac.uk}}
\author[2]{Alexandre Thiery\thanks{a.h.thiery@nus.edu.sg}}
\author[3]{Anthony Lee \thanks{anthony.lee@warwick.ac.uk}}
\affil[1]{Department of Mathematics and Statistics, Lancaster University}
\affil[2]{Department of Statistics, National University of Singapore}
\affil[3]{Department of Statistics, University of Warwick}

\begin{document}

\maketitle

\begin{abstract}
We consider a pseudo-marginal Metropolis--Hastings kernel $P_m$ that is constructed using an average of $m$ exchangeable random variables, as well as an analogous kernel $P_s$ that averages $s<m$ of these same random variables. Using an embedding technique to facilitate comparisons, we show that the asymptotic variances of ergodic averages associated with $P_m$ are lower bounded in terms of those associated with $P_s$.
We show that the bound provided is tight and disprove a conjecture that when the random variables to be averaged are independent, the asymptotic variance under $P_m$ is never less than $s/m$ times the variance under $P_s$. The conjecture does, however, hold when considering continuous-time Markov chains. These results imply that if the computational cost of the algorithm is proportional to $m$, it is often better to set $m=1$. We provide intuition as to why these findings differ so markedly from recent results for pseudo-marginal kernels employing particle filter approximations. Our results are exemplified through two simulation studies; in the first the computational cost is effectively proportional to $m$ and in the second there is a considerable start-up cost at each iteration. 
\end{abstract}

\section{Introduction}
\label{sec.introduction}

The Metropolis--Hastings algorithm is often used to approximate expectations with respect to
posterior distributions, making use of  
point-wise evaluations of the posterior density $\pi$ up to a fixed but arbitrary constant of proportionality. In cases where such evaluations are infeasible, the pseudo-marginal Metropolis--Hastings algorithm \cite[][]{Beaumont03,AndrieuRoberts:2009} can be used if a realisation of a non-negative, unbiased stochastic
\textit{estimator} of the
target density, possibly up to an unknown normalisation constant, is
available. These estimators can, for example, be constructed using
importance sampling \citep{Beaumont03}, 
a particle filter or sequential Monte Carlo \cite[]{AndrieuDoucetHolenstein:2010}. 

A key tuning parameter of such pseudo-marginal algorithms is the 
number of samples or particles, which we denote by $m$, 
and we are interested in the relationship
between $m$ and the computational efficiency of the pseudo-marginal algorithm for approximating posterior expectations. The algorithm is a type of Markov chain Monte Carlo method: a Markov chain with stationary distribution $\pi$ is simulated for a finite number of steps in order to compute an appropriately normalised partial sum. This quantity then serves as an approximation of a limiting ergodic average that is almost surely equal to the expectation of interest. One natural measure of computational inefficiency, defined precisely in the sequel, is the asymptotic variance of the ergodic average of interest multiplied by the computational effort required to simulate each value of the Markov chain.
\cite{AndrieuVihola:2015} showed that
the asymptotic variance for a pseudo-marginal algorithm is bounded below by
the asymptotic variance of an idealised algorithm in which $\pi$ 
is evaluated exactly, so one can think of the relative asymptotic variance of a pseudo-marginal ergodic average as its asymptotic variance divided by the asymptotic variance of the idealised ergodic average. In some senses, this idealised algorithm is
approached as $m\uparrow\infty$ and one might suppose that the relative asymptotic variance
should therefore decrease to $1$ as $m$ increases. This is indeed true for
estimators that arise from importance sampling \citep{AndrieuVihola:2014}, at least when the asymptotic variance is finite for some finite $m$. An important issue, then, is whether the decrease in asymptotic variance at the expense of increased computational effort is justified in terms of computational efficiency. 

In this article, we consider arbitrary pseudo-marginal algorithms
where the posterior density is estimated using an average of unbiased
estimators, such as with importance sampling, and we show
that in all such cases the asymptotic variance when $m$ samples are
used is not much smaller than the asymptotic variance when a single
sample is used, divided by $m$. Thus if the computational cost is roughly proportional to $m$, as is often the case, there is little, if any, gain in using more than one sample. This formalises empirical observations made in Section~3.4 of an early version of \cite{STRR:2015} (available at http://arxiv.org/abs/1309.7209v1), and generalises the main result of \cite{BPSW2016}, which assumes that the pseudo-marginal kernels are positive and the estimators to be averaged are independent and take only one non-zero value. We demonstrate that our bound is tight and illustrate it  through two
simulation studies. In the second study, an additional fixed and large cost is associated with simulating $m$ samples,  so that the
computational efficiency is maximised 
at some $m \gg 1$. Our result also implies that asymptotic variance being infinite for $m=1$ implies that it is infinite for all finite $m$.
The theory  also suggests that when parallel computation is feasible, it should be close to optimal to obtain one estimate on each core and then to average these. We discuss why this result differs so markedly from the particle-filter scenario, which was analyzed under significantly stronger conditions than here by \cite{STRR:2015} and \cite{DPDK2015}.

We adopt the notation $x \wedge y=\min\{x,y\}$; for an integer $m \geq 0$, we set $[m] \equiv \{1,2, \ldots, m\}$. For a probability measure $\pi$ on some measurable space $(\sX,\Sigma)$ and a $\pi$-integrable test function $\phi:\sX \to \bR$, we define  $\pi(\phi) \equiv \int_{\sX} \phi(x) \, \pi(\md x)$ and use the notation $L^2(\pi) \equiv \curBK{ \phi: \mathsf{X} \to \bR \, : \, \pi(\phi^2) < \infty}$ to  designate the usual Hilbert space with norm $\norm{\phi}_\pi^2 = 
\pi(\phi^2)$.

\section{Main results}
\label{sect.setup.results}
\subsection{Asymptotic variance of ergodic averages and accept-reject kernels}
\label{sect.ineff.def}
Consider a Markov transition kernel $P$ with invariant distribution $\pi$ and associated Markov chain $\{X_k \}_{k=0}^\infty$ with $X_0 \sim \pi$. For any $\phi \in L^2(\pi)$, an estimator of $\pi(\phi)$ is the ergodic average $n^{-1}\sum^n_{k=1}\phi(X_k)$ and the asymptotic variance of the ergodic average is 
\begin{align*}
\lVar(\phi,P)
\equiv 
\lim_{n \to \infty} \, \mbox{var}\left\{n^{-1/2} \, \sum_{k=1}^n \phi(X_k)\right\}.
\end{align*}
%
All Markov kernels $P$ in this article involve accepting or rejecting a sample from a proposal kernel $Q$
according to an acceptance probability $\alpha(x;x')$. We define the
marginal acceptance probability from $x$, $\alpha(x)=\int_{\sX}\alpha(x;x') \, Q(x,\md x')$.
The kernel $P$ is then of the form
\begin{align}
\label{eqn.acc.rej.gen}
P(x, \md x')
\, \equiv \, 
\{ 1-\alpha(x)\} \, \delta_x(\md x')
\, + \, 
Q(x,\md x') \, \alpha(x;x').
\end{align}

\subsection{Pseudo-marginal Metropolis--Hastings}
\label{sect.PMMH}
Let $\pi(\md x) = \pi(x) \, \md x$  be a probability distribution on $\sX$, where $\md x$ denotes a dominating measure,
and $Q$ be a proposal kernel with density $q$, i.e.  $Q(x,\md x')=q(x,x') \, \md x'$. The $\pi$-reversible Metropolis--Hastings kernel associated with $Q$ is defined via \eqref{eqn.acc.rej.gen} by
taking the acceptance probability
$\alpha(x;x')\equiv1\wedge r(x,x')$ where $r(x,x')$ is the 
Metropolis--Hastings ratio,
\begin{align*}
r(x,x')
\equiv
\frac{\pi(x') \, q(x',x)}{\pi(x) \, q(x,x')}.
\end{align*}

When pointwise relative evaluation of the density $\pi$ is not possible, $\alpha(x;x')$ is intractable. The pseudo-marginal Metropolis--Hastings algorithm introduces an unbiased approximation to the posterior, $\pihat(x,U)$, where $U$ is a vector of
auxiliary random variables. We will be interested in the random
variable $W \equiv \pihat(x,U)/\pi(x) \in \sW \subseteq [0,\infty)$ which satisfies $\Expect{W}=1$. The pseudo-marginal algorithm simulates a Metropolis--Hastings Markov chain on the extended state space $\sX \times \sW$ with proposal density $q(x, x') \, q_{x'}(w')$ and invariant density $\pitil(x,w)  =  \pi(x)  q_{x}(w)  w$,
i.e. proposals are accepted with the usual Metropolis--Hastings ratio, which in this case reads $\alpha(x,w;x',w'):=1 \wedge \{ r(x,x') \, w'/w \}$. Importantly, $\pitil$ admits $\pi$ as its $x$-marginal.

\subsection{Pseudo-marginal algorithm using averages}
\label{sect.PMMHimport}
Suppose that for each $x \in \sX$ it is possible to generate an
unbiased non-negative estimator
$\pi(x) \, W$
of the target
density $\pi(x)$, i.e. $\Expect{W} = 1$ for any $x\in  \sX$. For any integer $r \geq 1$, one may use an
average of $r$ such estimators to construct the unbiased estimator
$\pi(x) \, \BK{W_1 + \ldots + W_r}/r$. In what follows, we assume $\ulW = (W_1, \ldots, W_r) \in \sW^{r}$
is exchangeable with joint density $q_x(\ulw)$. This accommodates the scenario  where $W_1, \ldots, W_r$ are independent and distributed according to $q_x(w)$ so $q_x(\ulw)=q_x(w_1)\cdots q_x(w_r)$.
We denote the associated kernel, acceptance probability and marginal acceptance probability 
by 
$\PPM_{r}$,
$\alpha_{r}(x,\ulw;x',\ulw')$ and 
$\alpha_{r}(x,\ulw)$, respectively.

Corollary $31$ of \cite{AndrieuVihola:2014} shows that, for two positive integers $s \leq m$, asymptotic variances associated with $\PPM_{m}$ are
at most those associated with $\PPM_{s}$ for $L^2(\pi)$ functions of the $x$-coordinate only.
Given this ordering, it is natural to ask
whether the decrease in asymptotic variance is sufficient to justify the extra computational
expense of $\PPM_{m}$. \cite{AndrieuVihola:2015} show that the 
asymptotic variance of a pseudo-marginal algorithm ergodic average is bounded below by that of the algorithm in which $\pi$ is evaluated exactly; consequently, there must eventually be diminishing returns for any increase in $m$. 
For functions $\phi \in L^2(\pi)$ of the $x$-coordinate only we are interested in $\mbox{var}\{n^{-1}\sum_{k=1}^n\phi(X_k)\}\approx n^{-1}\lVar(\phi,\PPM_r)$.  
A reduction in variance equivalent to that obtained by increasing $r$ from $s$ to $m$ could instead be obtained by increasing $n$ by a factor of $\lVar(\phi,\PPM_s)/\lVar(\phi,\PPM_m)$. Since computational time is proportional to $n$, 
if it is also proportional to the number of samples per iteration, $r$, a natural way of comparing the two Markov kernels $\PPM_{s}$ and $\PPM_{m}$ is through their computational inefficiencies $s \lVar(\phi, \PPM_{s})$ and $m \lVar(\phi, \PPM_{m})$, respectively.
As shown in the sequel, these quantities are not ordered in general but Theorem \ref{thrm.main} below shows that the quantities $r \curBK{ \lVar(\phi, \PPM_{r}) + \Vars{\pi}{\phi}}$ are. Since in many situations $\Vars{\pi}{\phi} \ll \lVar(\phi, \PPM_{r})$ this can be viewed as almost ordering computational inefficiencies.

\begin{theorem}
\label{thrm.main}
For positive integers $s \leq m$, the pseudo-marginal kernels $\PPM_{s}$ and $\PPM_{m}$ satisfy
\begin{align} \label{eq.main.var.ineq}
s \{ \lVar(\phi, \PPM_{s}) + \Vars{\pi}{\phi}\}
\; \leq \;
m \{ \lVar(\phi, \PPM_{m}) + \Vars{\pi}{\phi}\},
\end{align}
for any function $\phi \in L^2(\pi)$ of the $x$-coordinate only.
\end{theorem}
\begin{remark}
The inequality \eqref{eq.accept.ineq} in the proof also implies through fairly simple manipulations that the average acceptance rates satisfy $\alpha_m \le (m/s) \alpha_s$.
\end{remark}
One interesting consequence is that the class of $L^2(\pi)$ functions with finite asymptotic variance, which is often not all of  $L^2(\pi)$ \citep{lee2014variance}, cannot be enlarged by increasing $m$.
\begin{corollary}
In combination with Corollary~31 of \cite{AndrieuVihola:2014} we obtain that for $\phi \in L^2(\pi)$, $\lVar(\phi, \PPM_{m}) < \infty \iff \lVar(\phi, \PPM_{1}) < \infty$.
\end{corollary}

For a positive $\pi$-reversible Markov kernel $P$, $\lVar(\phi,P) \geq
\Vars{\pi}{\phi}$ for all $\phi \in L^2(\pi)$. Consequently,
Theorem~\ref{thrm.main} leads to the following generalisation of
Proposition~$4$ of \cite{BPSW2016}. 
\begin{corollary}
\label{cor.extpillai}
Let $s \leq m$ be positive integers and Markov kernel $\PPM_m$ be positive. For any function $\phi \in L^2(\pi)$ of the $x$-coordinate only, $\lVar(\phi, \PPM_s) \leq (2m/s - 1) \, \lVar(\phi, \PPM_m)$.
\end{corollary}
Positivity of 
 some random-walk-based kernels can be verified via results of \cite{DPDK2015} and \cite{Sherlock:2015}, which build upon \cite{baxendale2005renewal}. Independent Metropolis--Hastings pseudo-marginal kernels are always positive, as they are themselves independent Metropolis--Hastings kernels \citep{AndrieuVihola:2015}.

\subsection{Tightness of the result} \label{sec.counter-example}
The following proposition shows that the inequality in Theorem~\ref{thrm.main} cannot be improved in general, and that even if we consider averages of independent estimators the conjecture that $s \, \lVar(\phi, \PPM_{s}) \leq m \, \lVar(\phi, \PPM_{m})$ is not true in general.
\begin{proposition}\label{prop:tightness}
There exist pseudo-marginal kernels and $\phi \in L^2(\pi)$ such that
\begin{enumerate}
\item With negatively correlated $\ulW$, $\lVar(\phi,P_1)+\Vars{\pi}{\phi}=2\left\{\lVar(\phi,P_2)+\Vars{\pi}{\phi}\right\}$.
\item With independent $\ulW$, $\lVar(\varphi,P_{1}) > 2\lVar(\varphi,P_{2})$.
\end{enumerate}
\end{proposition}
The conjectured inequality, however, does hold in continuous time. Let $r \geq 1$ be an integer.
We define the continuous-time Markov chain, with kernel $\tilPPM_r$,
as the Markov chain whose transitions are identical to those of the discrete-time kernel $\PPM_r$ but take place on a Poisson clock with unit rate. That is, if $\tilX_r(t)$ is the $x$-process of a continuous-time Markov chain with transition $\tilPPM_r$ and $X_r(k)$ is the $x$-process of the discrete-time Markov chain with kernel $\PPM_r$, then $\tilX_r(t) = X_r(\mathrm{PP}[t])$ where $\{ \mathrm{PP}[t] \}_{t \geq 0}$ designates a Poisson process with unit rate. For $\phi \in L^2(\pi)$, the continuous-time asymptotic variance is defined as
\begin{align*}
\lVarCont(\phi, \PPM_r) = \lim_{T \to \infty} \; \frac{1}{T} ~ \Var{ \int_{0}^T \phi\BK{\tilX_r(t)} \, dt }.
\end{align*}
\begin{proposition}\label{prop:cts_version}
For positive integers $s \leq m$, the continuous-time chains satisfy
\begin{align*}
s \lVarCont(\phi, \PPM_s) \; \leq \; m \lVarCont(\phi, \PPM_m),\qquad \phi \in L^2(\pi).
\end{align*}
\end{proposition}

\section{Numerical Studies}
\label{sect.numerical}
\subsection{Preliminaries}
We present in this Section two numerical studies. Several choices of
proposal distributions are investigated and the situation when the
computational time necessary to generate $m$ samples is not
proportional to $m$, due to non-negligible computational overhead, is
carefully examined. Computational efficiency is measured in terms of
Effective Sample Size per unit of computational time; if the
computational time to generate $m$ samples in each of $n$ iterations is exactly proportional to $nm$, then we define $\ess(\phi,P) = n\Vars{\pi}{\phi} / \{\lVar(\phi,P)\}$ and $\ess^*(\phi,P) = \ess(\phi,P)/(nm)$. Since $\ess(\phi,P)$ and $\ess^*(\phi,P)$ are intractable in general, we consistently estimate it below using realisations of the Markov chain with kernel $P$.

\subsection{Inverse Stochastic Heat Equation}
We consider in this section, the problem of reconstructing the initial temperature field $u(t,x)$ at time $t=0$ and $x\in (0,1)$ from $N \geq 1$ noisy measurements at time $t=T$ distributed as $y_i = u(T, x_i) + \xi_i$ for some locations $\{x_i\}_{i=1}^N$ and independent centred Gaussian samples $\{\xi_i\}_{i=1}^N$ with variance $\sigma^2_{\xi}$. The temperature fields evolves according to the stochastic Heat equation 
\begin{align*}
\partial_t u = \Delta u + \sigma \, \dot{W}
\end{align*}
with Dirichlet boundary $u(t,0)=u(t,1)=0$, and where $\dot{W}$ is a
space-time white noise; for the simulations, we chose $\sigma =
10^{-2}$ and $T=2 \cdot 10^{-2}$. {\emph A priori}, we use a truncated Karhunen--Lo\`eve expansion to model $u(0,\cdot)$, i.e.
$u(0,x) = \sum_{k=0}^M \zeta_k \, \sin(k \, \pi \, x)$, where $\zeta_k \sim \mathcal{N}(0,k^{-2})$ are independent.
Simulations were carried
out by standard forward discretization of the stochastic partial
differential equations.
The pseudo-marginal algorithms are started in a region of high posterior mass; we used a Crank--Nicholson proposal of the type $u_\star = \alpha \, u + (1-\alpha^2)^{1/2} \, \zeta$, where $\zeta$ is distributed according to the prior distribution, with a value of $\alpha \in (-1,1)$ chosen such that the acceptance rate when $m \gg 1$ is around $1/2$. 
The computational efficiency is taken as the minimum computational efficiency associated with $9$ functions $u \mapsto u(0,i/10)$ for $i \in \{1,\ldots,9\}$.
Table \ref{tablelabel} shows approximate $90$\% confidence intervals and, as expected, the computational efficiency is maximized for $m=1$.

\begin{table}
\caption{Computational efficiency $(\ess^* \times 10^3)$: $90$\% confidence intervals} 
\begin{tabular}{|c|c|c|c|c|c|c|c|c|c|c|}
$m$ & 1 & 2 & 3 & 4 & 5 & 6 & 7 & 8 & 9 & 10 \\
CE & $(8.3, 10)$ & $(4,4.8)$ & $(2.7,3.2)$
& $(2,2.4)$ & $(1.6,2)$ & $(1.4,1.6)$ & $(1.2,1.4)$ & $(1,1.2)$
& $(.9,1.1)$ & $(.8,1)$ \\
\end{tabular}
\label{tablelabel}
\end{table}

\subsection{Logistic regression using a latent Gaussian process}
\label{sect.GP}
We consider a logistic regression model with fixed effects and a latent
Gaussian process, following exactly the approach of \cite{Sherlock:2015}. The likelihood function is
approximated by importance sampling with a data-dependent proposal distribution, similar to \cite{FilipponeGirolami:2014} and \cite{GiorgiDiggle:2015}.
The parameter space has dimension $6$ and the latent Gaussian process is required
at $L=144$ observation points. Whatever the value of $m$, at each
iteration, creation of the importance sampling
proposal involves a single $\mathcal{O}(L^3)$ Cholesky decomposition of an $L \times L$
matrix; each importance sample then costs
$\mathcal{O}(L^2)$. For small values of $m$ the start-up cost dominates the cost of simulating $m$ importance samples.

The posterior mean and covariance
matrix of the parameters were estimated from a
trial run. For the random-walk pseudo-marginal Metropolis algorithm with $m=100$, an approximately
optimal scaling of $\lambda=0.9$ was found. This scaling was then used
in all runs recorded since it should also
be approximately optimal for other values of $m$
\cite[]{Sherlock:2015}, and thus help to control the Monte Carlo variability
in empirical effective sample sizes.  A set $\mathcal{M}\equiv\{1,2,3,4,5,10,20,40,100,200,400,1000,2000,4000\}$ of numbers of
importance samples was considered. For $m\in\{1,2,3,4,5,10,20,40,100,200,400\}$, the
pseudo-marginal algorithm was run for $2 \times 10^6$ iterations and
thinned for storage by a factor of $20$. For larger $m$ values the
run-lengths and thinning factors were respectively $10^6$ and $10$
($m=1000$), $4\times 10^5$ and $5$ ($m=2000$), and $2\times 10^5$ and
$2$ ($m=4000$). These run lengths kept the CPU time for each run
between $10^5$ and $3\times 10^5$ seconds and the effective sample sizes above
$500$. Despite the large run lengths, the Monte Carlo variability was
non-negligible for $m\le 5$ and so three independent runs were
performed for each of these $m$ values.

\begin{figure}[h]
\begin{center}
  \includegraphics[scale=0.3,angle=0]{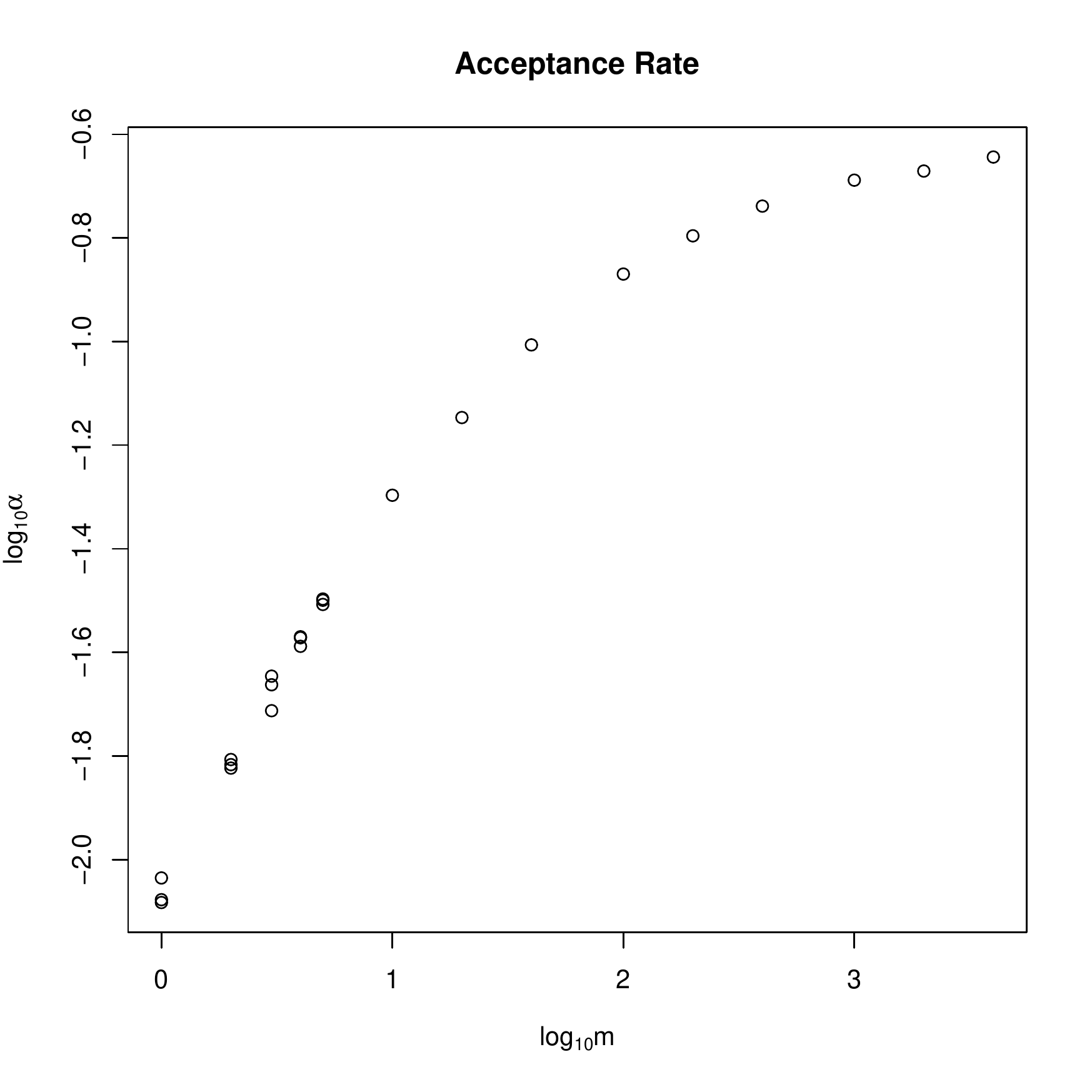}
  \includegraphics[scale=0.3,angle=0]{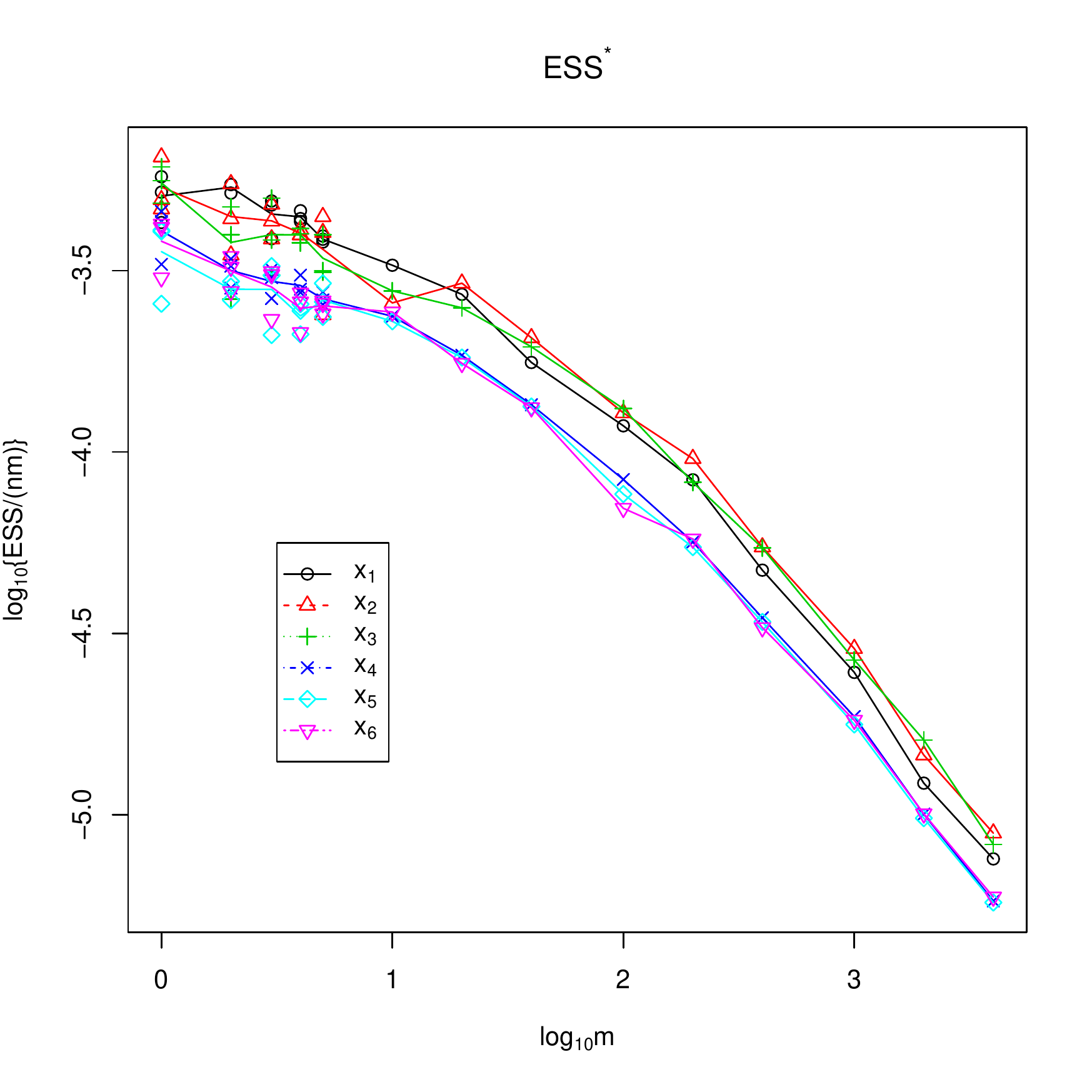}
  \includegraphics[scale=0.3,angle=0]{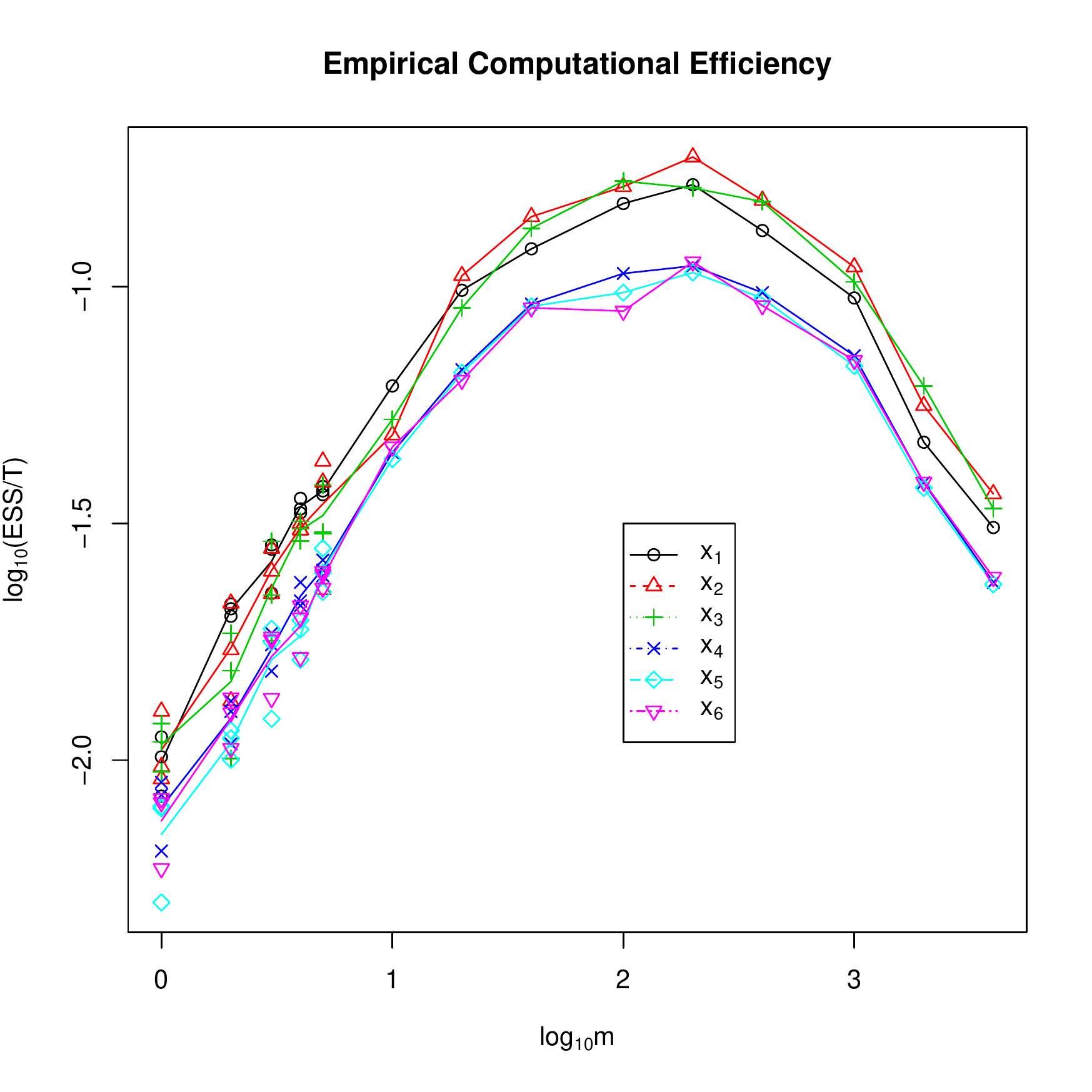}
\end{center}
\caption{
Log quantities for each run
against $\log_{10} m$ for
(left) acceptance rate, (centre) hypothetical computational efficiency
(ESS/$\{nm\}$) and (right) empirical computational efficiency (ESS/processing
time). Each line in the latter two graphs corresponds to one of the six parameters. 
\label{fig.simstudGP}
}

\end{figure}

Figure \ref{fig.simstudGP} reports the average acceptance rate, the
hypothetical computational efficiencies, $\ess^*$, and the
empirical computational efficiencies, $\ess$ divided by CPU time, for all six parameters. Due
to the non-negligible computational overhead, the two efficiency
measures are significantly different. A pseudo-marginal independence
sampler was also run and gave similar results. As the theory suggests, increasing from $s$ to $m$ samples never increases the acceptance rate by more than $m/s$ and 
the hypothetical computational efficiency is maximised at $m=1$. However, due to
the considerable start-up cost at each iteration, the empirical computational
efficiency is maximised at around $m=200$. Interestingly, it is at
$m=200$ that the cost of creating $m$ samples approximately matches
the start-up cost.

\section{Averaging versus particle filtering}
\label{sect.discuss.difference}
We have shown that if the computational cost of obtaining $m$ estimators is proportional to $m$ then
it is close to optimal to choose $m=1$ when averaging, at least when the asymptotic variance is finite. This is very different to \cite{STRR:2015} and \cite{DPDK2015}, who found under specific assumptions that when the likelihood function of a large
number of observations is estimated via a particle filter, $m$ should be chosen so that the variance of the
the log-likelihood estimator is controlled: the optimal choice of $m$ is consequently typically greater than one.

This fundamental difference arises because an estimator obtained using a particle filter is not an average, but a product of $T$ dependent averages of $m$ random variables. The relative variance of an importance sampling estimator with $m$ samples is $C/m$ for some $C>0$, whereas the relative variance of a particle filter estimate of the likelihood is of the form  \citep{cerou2011nonasymptotic},
$$\sum^T_{r=1} \left(\frac{1}{m} \right)^r \left(1-\frac{1}{m} \right)^{T-r} C_r,$$ where $C_1,\ldots,C_T$ is 
 is a non-negative sequence that often increases exponentially. It follows that increasing $m$ when $m$ is small can dramatically reduce the contributions of $C_2,\ldots,C_T$, even though by considering $m$ very large with $T$ fixed, the relative variance is $\mathcal{O}(C_1/m)$.

\bibliographystyle{apalike}
\bibliography{pmimport}

\appendix

\section*{Appendix}
If $P$ is reversible with respect to $\pi$, it can also be regarded as a self-adjoint operator on $L^2(\pi)$; the Dirichlet form associated with $P$ is defined as
\begin{align*}
\Dirichlet_P(\phi)\equiv\frac{1}{2}\int \pi(\md x)Q(x,\md x')\{\phi(x')-\phi(x)\}^2.
\end{align*}

\begin{proof} (of Theorem~\ref{thrm.main})
Recall Lemma $33$ of \cite{AndrieuLeeVihola:2015}: if for two $\mu$-reversible Markov
kernels $\Pi_1$ and $\Pi_2$ there exists $\rho>0$ satisfying
$\Dirichlet_{\Pi_2}(\phi) \geq \rho \, \Dirichlet_{\Pi_1}(\phi)$ for
all $\phi \in L^2(\mu)$, then
\begin{align}
\label{eqn.ALVDirIneq}
\rho\{\lVar(\phi, \Pi_2)+\Vars{\mu}{\phi}\} 
\; \leq \; 
  \lVar(\phi, \Pi_1)+  \Vars{\mu}{\phi}, \qquad \phi \in L^2(\mu).
\end{align}
To exploit this result, we construct two $\overline{\pi}$-reversible Markov kernels $\Pbar_s$ and $\Pbar_m$ on the extended space $\sX \times \sW^m \times [m]$ such that the $x$-marginal of the Markov chain with transition $\Pbar_s$ (resp. $\Pbar_m$) has the same law as the $x$-marginal of the Markov chain with transition $\PPM_s$ (resp. $\PPM_m$). By Lemma $33$ of \cite{AndrieuLeeVihola:2015}, Theorem \ref{thrm.main} follows once it is proved that for any $\phi \in L^2(\overline{\pi})$ it holds that
\begin{align} \label{eq.dirichlet.form.ineq}
\Dirichlet_{\Pbar_{m}}(\phi)
\; \leq \;
\frac{m}{s} \, \Dirichlet_{\Pbar_{s}}(\phi).
\end{align}
We define the distribution $\overline{\pi}$, which depends on $s$ and $m$, through its density
\begin{align} \label{eq.joint.target.cor}
\overline{\pi}(x, \ulw, k) 
&\equiv 
\frac{1}{m} \, \pi(x) \, \curBK{ \frac{w_k + \ldots + w_{k+s-1}}{s} } \, q_x(\ulw) 
=
\frac{1}{m} \, \pi(x)  \, q_x(\ulw) \, \mathrm{A}(\ulw,k) 
\end{align}
for $(x, \ulw, k) \in \sX \times \sW^m \times [m]$;
the indices in \eqref{eq.joint.target.cor}
and henceforth are to be understood modulo $m$, and 
we have used the notation $\mathrm{A}(\ulw,k) = (w_k + \ldots + w_{k+s-1})/s$.
The Metropolis--Hastings kernel $\Pbar_{s}$ proposes a move $(x,\ulw, k) \mapsto (X',\ulW', K')$ by first generating $(X', \ulW') \sim q(x, dx') \, q_{x'}(d\ulw')$ and then choosing $K'$ uniformly at random in $[m]$, i.e.  the proposal density is
\begin{align*}
\overline{q}_{s}(x,\ulw, k; x',\ulw', k') 
\equiv q(x, x') \, q_{x'}(\ulw') \, (1/m).
\end{align*}
The proposed $(X', \ulW',K')$ is accepted with the usual Metropolis--Hastings probability
\begin{align} \label{eq.accept.Ps}
\overline{\alpha}_{s}(x,\ulw, k; x',\ulw', k') 
\, = \,
1 \wedge \sqBK{r(x,x') \, \frac{w'_{k'}  + \ldots + w'_{k'+s-1}}{w_{k}  + \ldots + w_{k+s-1}}}.
\end{align}
The Metropolis--Hastings kernel $\Pbar_{m}$ differs from $\Pbar_{s}$ in the way $K'$ is proposed. It proposes a move $(x,\ulw, k) \mapsto (X',\ulW', K')$ by first generating  $(X', \sW') \sim q(x, dx') \, q_{x'}(d\ulw')$ and then choosing $K' \in [m]$ such that $\Prob{K'=k'} \propto  A(\ulw', k')$. Since for any $\ulw \in \sW^m$ we have $\mathrm{A}(\ulw,1)+ \ldots + \mathrm{A}(\ulw,m) = w_1 + \ldots + w_m$, the proposal density is
\begin{align*}
\overline{q}_{m}(x,\ulw, k; x',\ulw', k') 
\equiv
q(x, x') \, q_{x'}(\ulw') \, \frac{\mathrm{A}(\ulw',k')}{w'_1 + \ldots + w'_m}.
\end{align*}
The proposed $(X', \ulW',K')$ is accepted with the usual Metropolis--Hastings probability
\begin{align} \label{eq.accept.Pm}
\overline{\alpha}_{m}(x,\ulw, k; x',\ulw', k') 
=
1 \wedge \sqBK{r(x,x') \, \frac{w'_1 + \ldots + w'_m}{w_1 + \ldots + w_m}}.
\end{align}
From Equations \eqref{eq.accept.Ps}--\eqref{eq.accept.Pm} it follows that the $x$-coordinates of the Markov chains with transition $\Pbar_{s}$ and $\Pbar_{m}$ equal in law, respectively, the $x$-coordinates of the Markov chains with transitions $P_{s}$ and $P_{m}$.
To conclude the proof, we now prove inequality \eqref{eq.dirichlet.form.ineq}; it suffices to prove that 
$\Pbar_{m}(x,\ulw,k; x',\ulw',k')$ is at most $(m/s) \, \Pbar_{s}(x,\ulw,k; x',\ulw',k')$ for any $(x,\ulw,k) \neq (x',\ulw',k')$, i.e.
\begin{align} \label{eq.accept.ineq}
\frac{\mathrm{A}(\ulw',k')}{w'_1 + \ldots + w'_m} \, 
\overline{\alpha}_{m}(x,\ulw, k; x',\ulw', k')
\leq
(m/s) \, \{ m^{-1} \, \overline{\alpha}_{s}(x,\ulw, k; x',\ulw', k') \}.
\end{align}
From \eqref{eq.accept.Ps}--\eqref{eq.accept.Pm}, this is equivalent to showing that $1 \wedge \sqBK{r(x,x') \, (w'_1 + \ldots + w'_m)/(w_1 + \ldots + w_m)}$ is at most
\begin{align*}
\curBK{\frac{w'_1 + \ldots + w'_m}{s \, A(\ulw',k')}} \wedge \sqBK{r(x,x') \, \frac{w'_1 + \ldots + w'_m}{w_1 + \ldots + w_m} \, \curBK{ \frac{w_1 + \ldots + w_m}{s \, A(\ulw,k)}}},
\end{align*}
and since the two quantities inside curly brackets are at least one, the conclusion follows.
\end{proof}

%
%
%

\begin{proof} (of Proposition~\ref{prop:tightness})//
For 1., let $\sW=\{0,2\}$, $m=2$, $s=1$ and $q_x(0,2)=q_x(2,0)=1/2$. Since $(W_1+W_2)/2=1$, for
any test function $\phi(x)$, $\lVar(\phi,P_2)=\lVar(\phi,P_{\rm MH})$,
where $P_{\rm MH}$ is the Metropolis-Hasting kernel from which the
pseudo-marginal kernels are derived. However,
since, at stationarity, $W_1=2$,
\begin{eqnarray*}
\Dirichlet_{P_1}(\phi)
&=&
\frac{1}{2}\int
\pi(\md x)q_x(\md w)w~Q(x,\md x')q_{x'}(\md w')\left\{1\wedge \frac{w_1'\pi(x')}{w_1\pi(x)}\right\}\{\phi(x')-\phi(x)\}^2\\
&=&\frac{1}{4}\int
\pi(\md x)~Q(x,\md x')\left\{1\wedge \frac{\pi(x')}{\pi(x)}\right\}\{\phi(x')-\phi(x)\}^2
=\frac{1}{2}\Dirichlet_{P_{\rm MH}}(\phi).
\end{eqnarray*}
Applying \eqref{eqn.ALVDirIneq} with $\varrho=1/2$
and again with $\varrho=2$ then provides
\[
\lVar(\phi,P_1)+\Vars{\pi}{\phi}=2\left\{\lVar(\phi,P_2)+\Vars{\pi}{\phi}\right\}.
\]
For 2., let $\sX=\{1,2\}$, $\sW=\{0,1,4\}$, $\pi(\{1\})=2\pi(\{2\})$, $Q(1,\{2\})=Q(2,\{1\})=1$, $q_{1}(\{1\})=1$ and $q_{2}(\{4\})=1/4=1-q_{2}(\{0\})$. Since the Markov chains considered are finite, asymptotic variances can be calculated exactly. In particular, for $P_1$ and $P_2$ the sets $\{(1,1),(2,4) \}$ and $\{(1,1),(2,2),(2,4)\}$ are respectively absorbing. The transition matrices for these states can be calculated respectively as 
$$\left(\begin{array}{cc}
3/4 & 1/4\\
1/2 & 1/2
\end{array}\right),\qquad \left(\begin{array}{ccc}
9/16 & 3/8 & 1/16\\
1 & 0 & 0\\
1/2 & 0 & 1/2
\end{array}\right),$$
and asymptotic variances can be computed, e.g., using the expressions on p. 84 of \cite{Kemeny1969}. With $\phi : \sX \rightarrow \mathbb{R}$ defined by $\phi(1)=-1/2$ and $\phi(2)=1$, we obtain $\lVar(\varphi,P_{1})=5/6$ and $\lVar(\varphi,P_{2})=1/3$, so $\lVar(\varphi,P_{1})>2\lVar(\varphi,P_{2})$.
\end{proof}

\begin{proof} (of Proposition~\ref{prop:cts_version})\\
For an integer $r \geq 1$, a lag $k \geq 0$ and a test function $\phi \in L^2(\pi)$, set $C_r(\phi; k) = \mathrm{cov}\BK{\phi[X_r(0)], \phi[X_r(k)]}$ the lag-$r$ auto-covariance at equilibrium of the kernel $\PPM_r$. Similarly, for a lag $u>0$, set $\widetilde{C}_r(\phi; u) = \mathrm{cov}\BK{\phi[\tilX_r(0)], \phi[\tilX_r(u)]}$ the lag-$u$ auto-covariance at equilibrium in continuous time of the kernel $\tilPPM_r$. Since $\lVar(\phi, \PPM_r)$ can also be expressed as $C_r(\phi;0) + 2 \, \sum_{k=1}^{\infty} C_r(\phi;k)$, standard manipulations yield that
\begin{align} \label{eq.link.continuous.discrete}
\begin{aligned}
\lVarCont(\phi, \PPM_r)
&=
2 \, \int_{0}^\infty \widetilde{C}_r(\phi;u) \, du
=
2 \, \int_{0}^\infty \curBK{ \sum_{k=0}^{\infty} e^{u} \, \frac{u^k}{k!} \, C_r(\phi;k) } \, du\\
&=
2 \, \sum_{k=0}^{\infty} C_r(\phi; k)
=
\BK{\Vars{\pi}{\phi} + \lVar(\phi, \PPM_r)}.
\end{aligned}
\end{align}
After rearranging terms and using \eqref{eq.link.continuous.discrete}, the inequality $s \lVarCont(\phi, \PPM_s) \leq m \lVarCont(\phi, \PPM_m)$ reduces to \eqref{eq.main.var.ineq}, proved in Theorem \ref{thrm.main}.
\end{proof}

\end{document}